\documentclass{PoS}

\usepackage{mathbbol}
\usepackage{bm}

\title{Improved interpolating fields for hadrons at non-vanishing momentum
}

\ShortTitle{Improved interpolating fields for hadrons at non-vanishing momentum
}

\author{M.~Della~Morte$^{1,2}$, B.~J\"ager$^{\dagger,1,2}$, T.D.~Rae\thanks{Supported by DFG grant HA4470/3-1}$~^{,\dagger,1}$, H.~Wittig$^{1,2}$ \phantom{\speaker{B.~J\"ager, T.D.~Rae}}\\
       
        $^{1}$ PRISMA Cluster of Excellence and Institut f\"ur Kernphysik, Becher-Weg 45, University of
	Mainz, D-55099 Mainz, Germany\\
	$^{2}$ Helmholtz Institute Mainz, University of Mainz, D-55099 Mainz, Germany\\
 	E-mail: \email{thrae@uni-mainz.de}\\
 	E-mail: \email{jaeger@kph.uni-mainz.de}}

\abstract{\vspace{-9.5cm} \phantom{a} \hfill HIM-2012-6 \newline \phantom{a}
\vspace{8.5cm} \newline We demonstrate that a reduction in the noise-to-signal ratio may be obtained for hadrons 
at non-zero momenta whilst maintaining a good overlap with the ground state through a generalisation 
of Gaussian/Wuppertal smearing. The use of an anisotropic smearing wavefunction is motivated by the
physical picture of a boosted hadron.
\newline

\begin{flushright}
\includegraphics[width=0.28\linewidth]{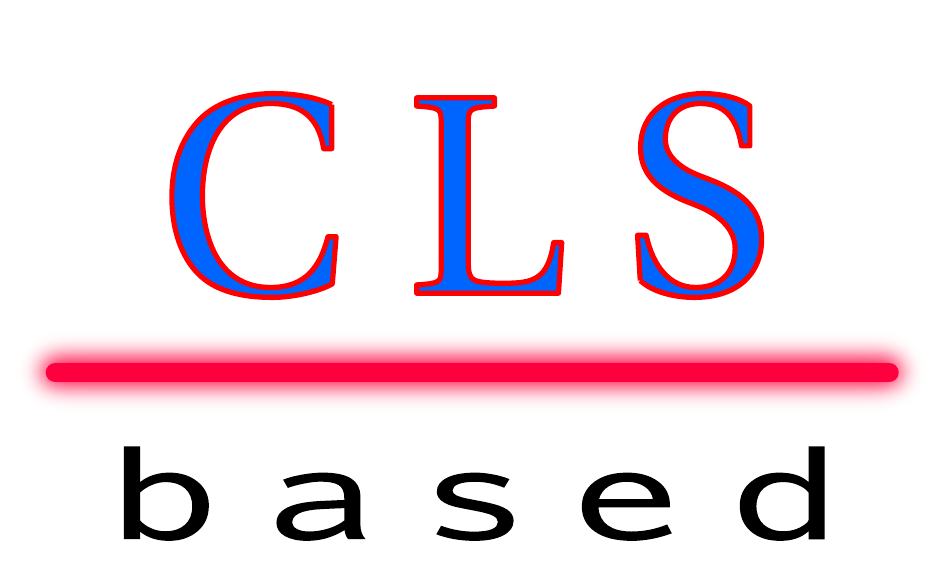}
\end{flushright}
}

\FullConference{The 30 International Symposium on Lattice Field Theory - Lattice 2012,\\
		June 24-29, 2012\\
		Cairns, Australia}

\begin{document}

\section{Introduction}
\begin{figure}
\centering
\includegraphics[width=0.62\linewidth]{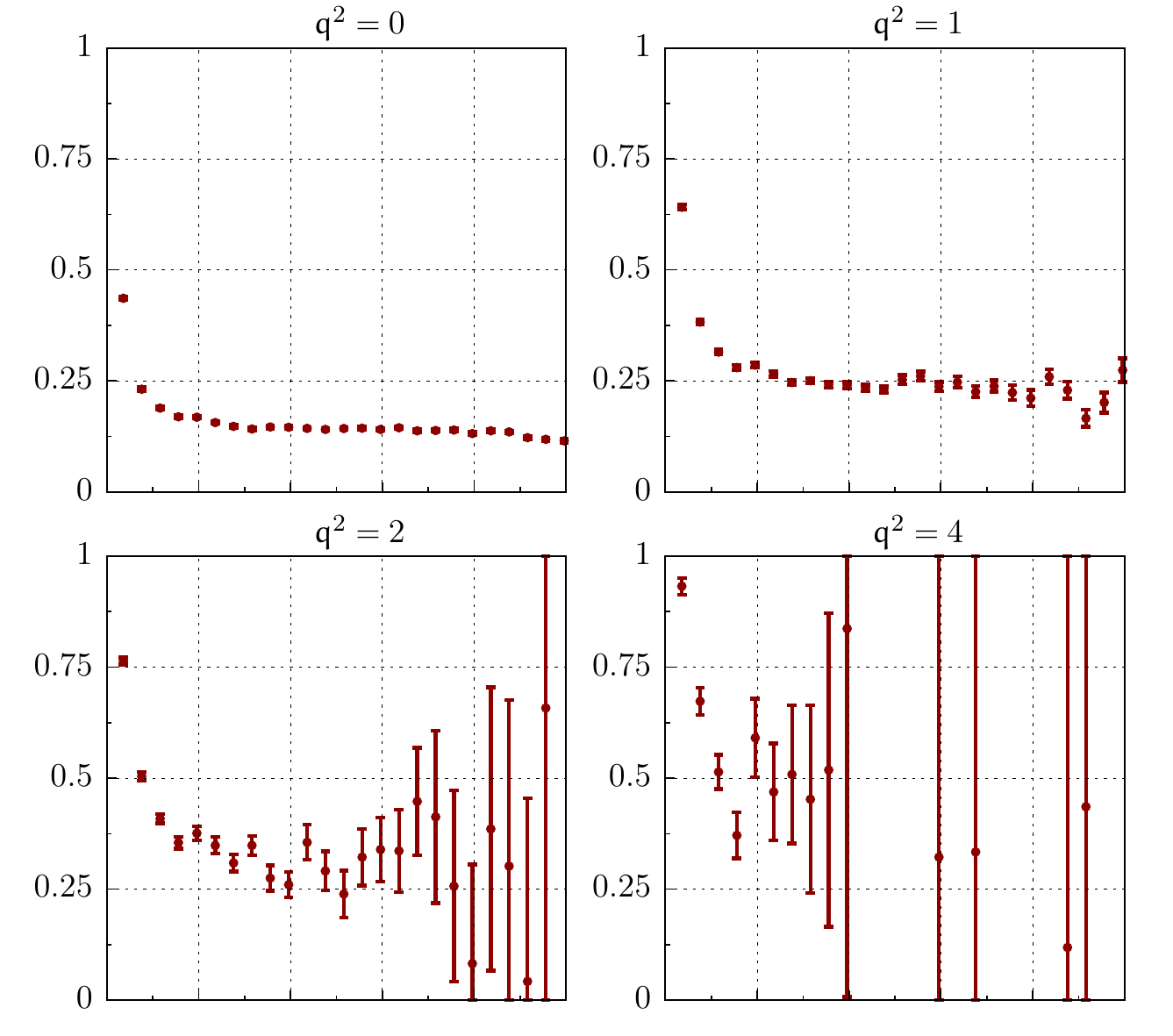}
\caption{\small Effective energies for pions at different values of $\bf{q}^2$.}
\label{fig:RNS}
\end{figure}

Lattice QCD is a useful tool for the extraction of quantities such as masses, energies and matrix 
elements. These may be extracted for a given hadron from the Euclidean time dependence of correlation 
functions for states with appropriate quantum numbers. The ground state may be isolated 
through taking the large source-sink separation limit due the time dependent exponential suppression 
of excited states. However the statistical errors of the correlation functions also grow with increasing 
time separations (with the exception of the pion at zero momentum), and so it can prove difficult to 
find a `window' where both statistical and systematic errors are under control. This becomes increasingly 
more difficult once states with non-zero 
momenta are considered, as is the case for many desirable quantities such as form factors. 
A clear example of the deterioration of the signal for increasing momentum $\bf{q}^2$ is given 
for the pion in figure~\ref{fig:RNS}. Throughout this paper we will use the notation 
${\bf p}={\bf q}~\frac{2\pi}{L}$ when referring to the momenta carried by the hadron, where 
${\bf q}=(q_1,q_2,q_3)$ is a vector of integers. 
For the case of the pion at zero momentum, the noise-to-signal ratio is expected to be constant as a 
function of time whereas for non-zero-momentum it is expected to be~\cite{Parisi,Lepage} 

\begin{equation}
R_{\rm NS}(x_0) \propto e^{(\sqrt{m_\pi^2+{\mathbf p}^2}-m_\pi) x_0}\;, \qquad {\rm for} \; x_0\to \infty \;.
\label{eq:RNSasy}
\end{equation}
This behaviour is demonstrated for our data in figures~\ref{fig:RNS} and \ref{fig:RNSasy} for several values of $\bf{q}^2$. 
The asymptotic trend sets in quite early and is consistent with noise dominated by a zero-momentum two 
pion state.

The usual technique used to improve the ground state overlap with the interpolating source fields, thus 
improving the `window', is through smearing. The present study is devoted to
optimising the smearing based 
upon the hadrons kinematics. In the following sections we detail how a generalisation of Gaussian/Wuppertal 
smearing that allows for anisotropic quark smearing (interpolating a boosted hadron) can lead to a 
significant improvement of the noise-to-signal ratio for states of non-zero momentum. 
The calculations were performed using $N_f=2$ non-perturbative $\mathcal{O}(a)$~improved Wilson fermions 
on configurations generated as part of the CLS initiative. This initial exploratory calculation utilises 
$168$ configurations, each with four evenly spaced source positions, on a lattice of $32^3\times 64$ points, 
with a lattice spacing of $0.063(2)~\textrm{fm}$ and a pion mass of approximately $450~\textrm{MeV}$; 
further details on the simulation parameters may be found in \cite{Fritzsch:2012wq,Capitani:2012gj,Capitani:2011fg}. 

\section{Implementation of anisotropic smearing}
\begin{figure}
\centering
\begin{minipage}[c]{0.5\linewidth}
    \includegraphics[width=1.0\linewidth]{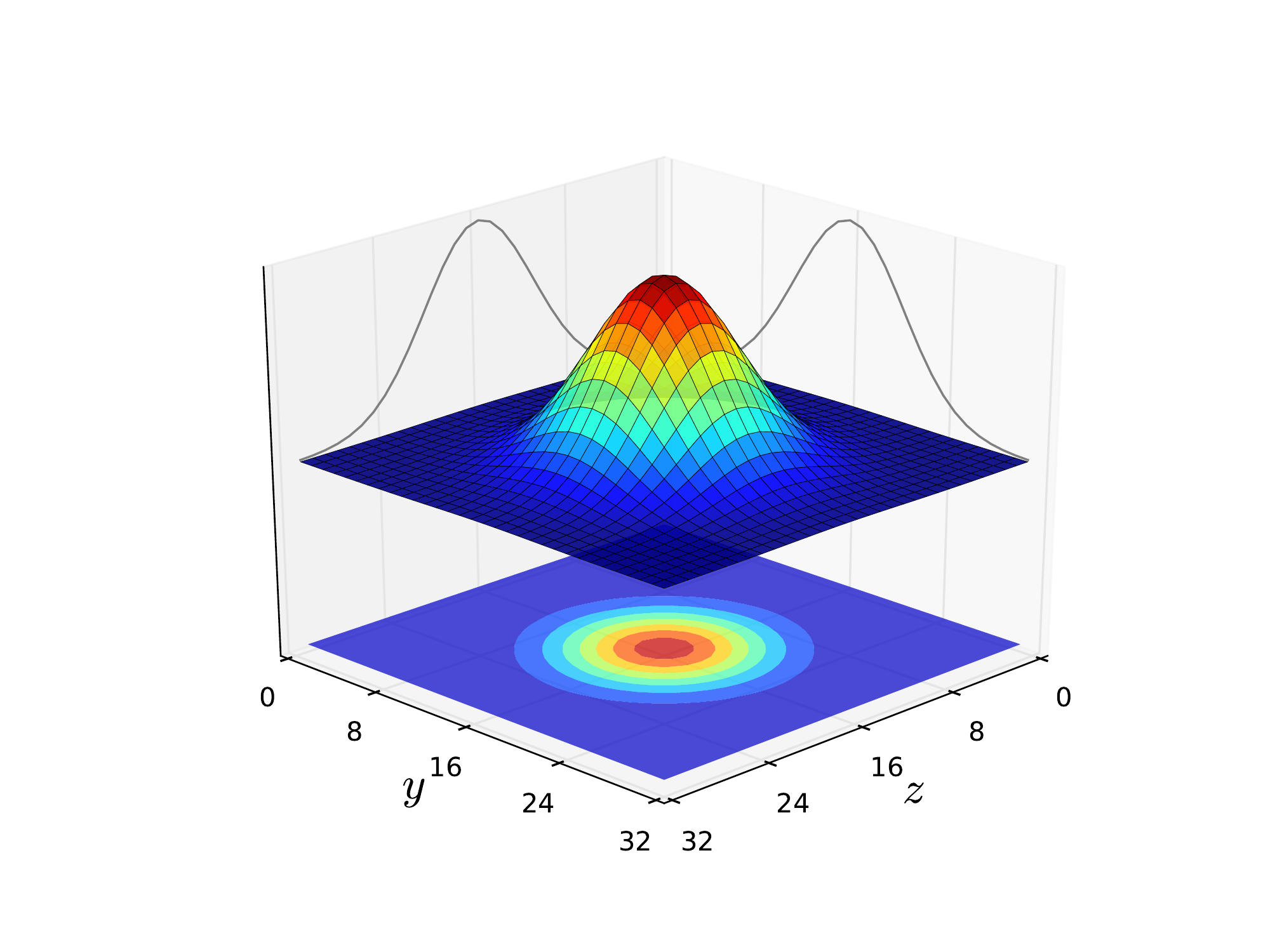}
  \end{minipage}
\hspace{-1cm}
  \begin{minipage}[c]{0.5\linewidth}
   \includegraphics[width=1.0\linewidth]{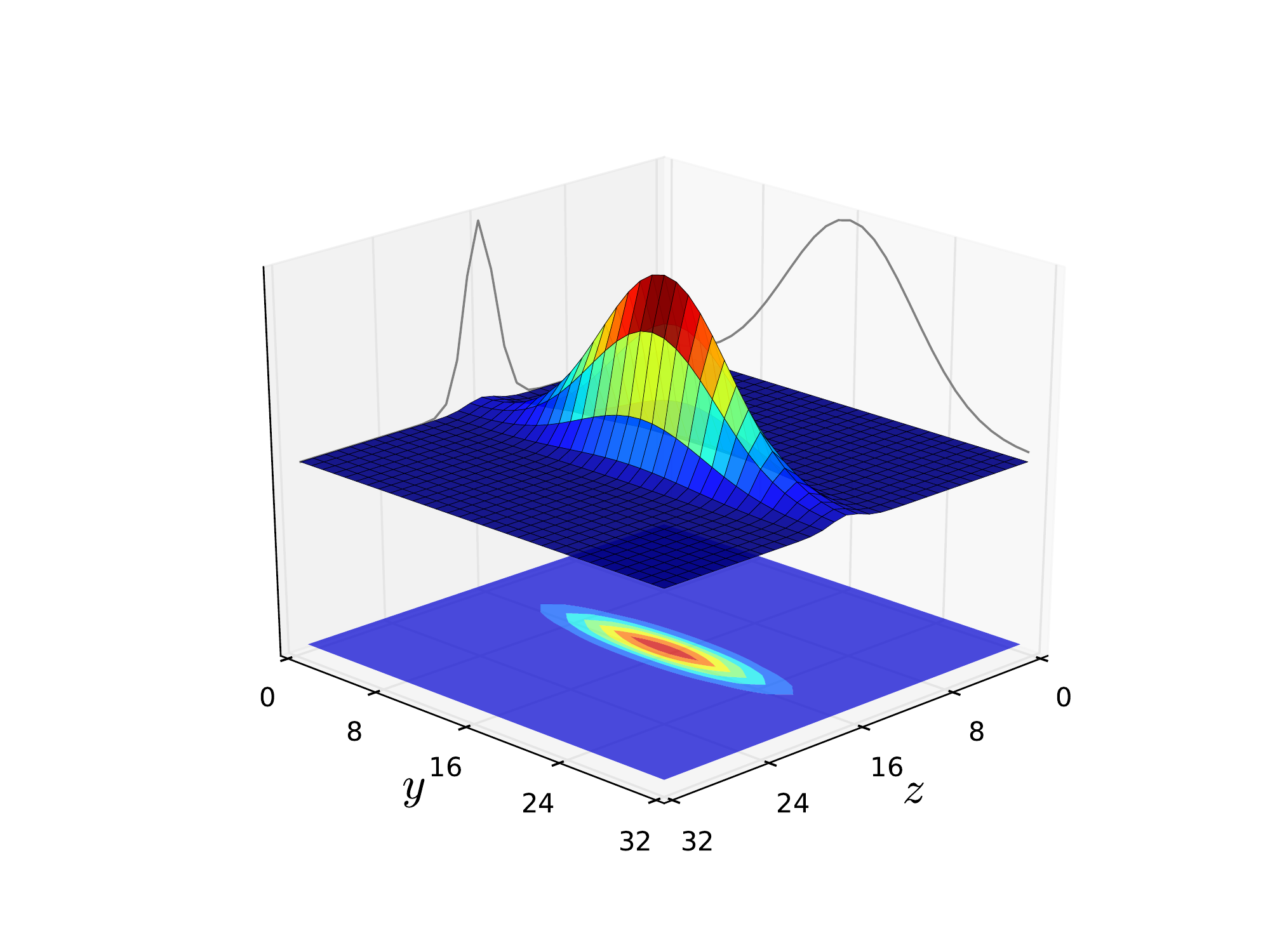}
  \end{minipage}
\caption{\small The $yz$-section of a spherical wavefunction $\kappa_x=\kappa_y=\kappa_z=2.9$ (left panel) and an anisotropic wavefunction $\kappa_x=\kappa_y=2.9,\kappa_z=2.9/16$ (right panel).}
\label{fig:wavefs}
\end{figure}
Gaussian/Wuppertal smearing is defined by~\cite{Gusken:1989ad,Gusken:1989qx}
\begin{equation}
H_k(x,x')= U_k(x) \delta_{x',x+\hat{k}}+U^\dagger_k(x-\hat{k}) \delta_{x',x-\hat{k}}\;,
\end{equation}
where $H_k$ is the hopping operator in direction $k$ and
the smeared fermion field $\psi^{(\kappa)}$ is defined as
\begin{equation}
\psi^{(\kappa)}(x) = (1+\kappa H)^n\, \psi(x)\;,
\label{eq:Wupp1}
\end{equation}
where 
\begin{equation}
H=H_x + H_y + H_z \;.
\label{eq:genH}
\end{equation}
In our generalization, we promote $\kappa$  and $H$ to spatial vectors $\boldsymbol{\kappa}$ and 
${\mathbf H}$ with components $x$, $y$ and $z$ so that
\begin{equation}
\psi^{(\boldsymbol{\kappa})}(x) = (1+\boldsymbol{\kappa}\cdot {{\mathbf H}})^n\, \psi(x)\;.
\label{eq:Wupp2}
\end{equation}
It is then possible to interpolate a boosted hadron through tuning the components in $\boldsymbol{\kappa}$ 
so that a Gaussian-like wavefunction is squeezed into something more point-like in the boosted direction. 
We also note that this tuning between point-like and gaussian-like enables a balance of parameters that 
result in, in the first case a better noise-to-signal ratio and in the latter a better overlap and thus plateau \cite{Lin:2010fv}.
Figure~\ref{fig:wavefs} shows the $yz$-cross sections corresponding to $\kappa_x=\kappa_y=\kappa_z=2.9$ 
(spherical - left panel) as well as  $\kappa_x=\kappa_y=2.9,\kappa_z=2.9/16$ (anisotropic - right panel). 
We use $\kappa_i=2.9$ as a basis and fix $n=140$ as these values maximised the length of the effective mass 
plateaus for the nucleon and pion at ${\bf q}^2=0$ and produce a wavefunction with a radius of approximately 
$0.5~\textrm{fm}$~\cite{Capitani:2012gj}. The anisotropic smearing is implemented so that only directions with momenta are squeezed. 
In addition we smear the gauge links in the hopping operator 
by applying one level of hypercubic smearing, using the parameters specified in the definition of ``HYP2''
links in \cite{DellaMorte:2005yc}.

\section{Pion and nucleon effective-mass study}  
\begin{figure}
\centering
\includegraphics[width=0.9\linewidth]{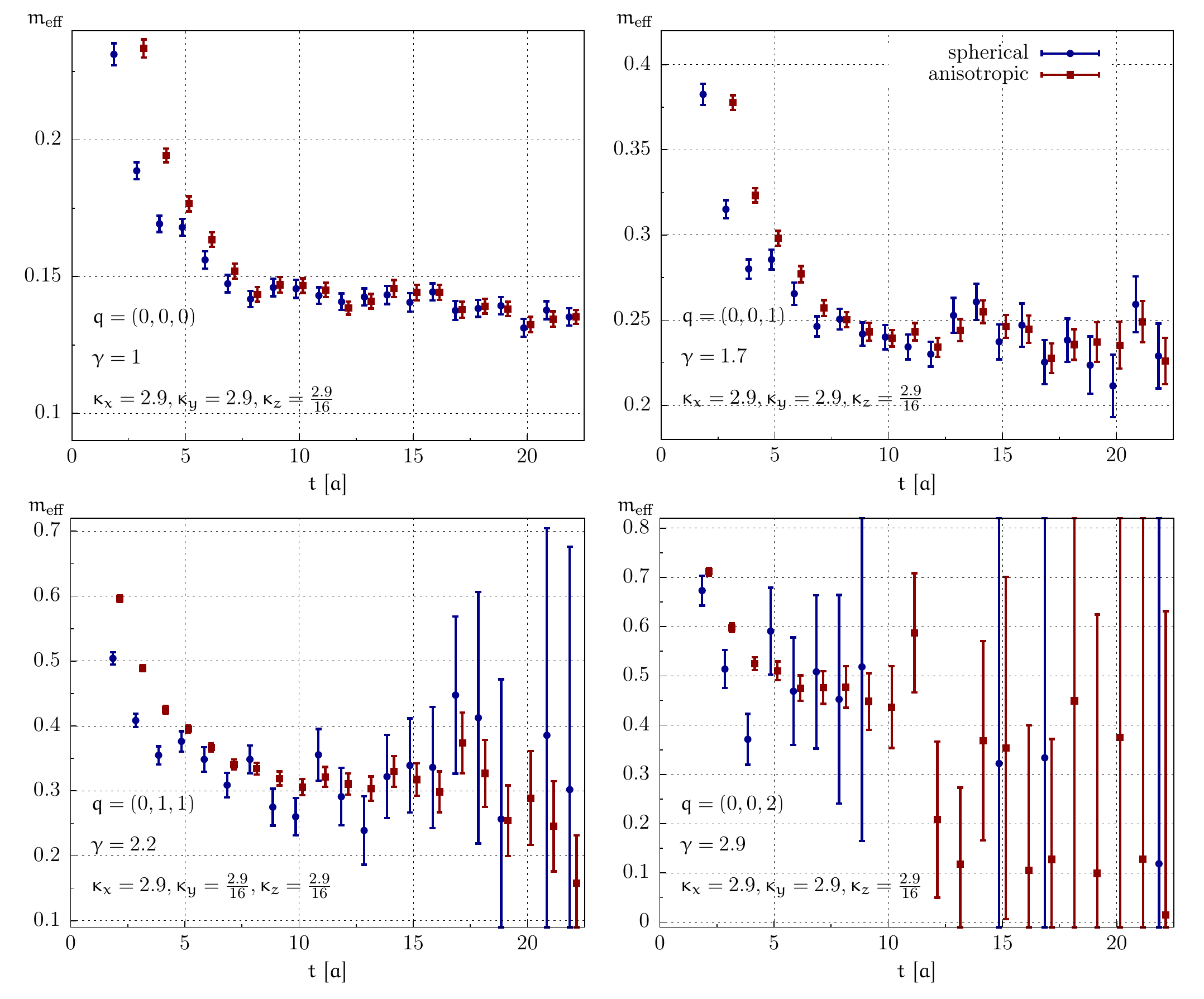}
\caption{\small Effective energies for the pion
 for different values of ${\bf q^2}$ using spherical (blue) and anisotropic 
(red) smearing.}
\label{fig:Meff_Pion}
\end{figure}

We consider two-point correlation functions of the form eq.~(\ref{eq:2pt}), where the operator $O$
interpolates mesons or baryons
\begin{equation}
C(x)= \left\langle O(\psi^{(\boldsymbol{\kappa})}(x),\overline{\psi}^{(\boldsymbol{\kappa})}(x)) \, 
O^\dagger(\psi^{(\boldsymbol{\kappa})}(0),\overline{\psi}^{(\boldsymbol{\kappa})}(0))\right\rangle \;.
\label{eq:2pt}
\end{equation}
Through the use of the anisotropic smearing we aim to reduce the contribution to the noise from the 
zero-momentum component of the four-point correlation function occurring in the variance of the two-point 
correlation function. Figure~\ref{fig:Meff_Pion} demonstrates the effect of anisotropic smearing for 
four values of $\bf{q}^2$. For each direction that the squeezing is applied to, the boost factor $\gamma$,
defined by
\begin{equation}
\gamma = {{\sqrt{m_{\rm H}^2 +{\bf p}^2}}\over{m_{\rm H}}},
\label{eq:Relgam}
\end{equation}
 is indicated.
\begin{figure}
\begin{center}
\includegraphics[width=0.62\linewidth]{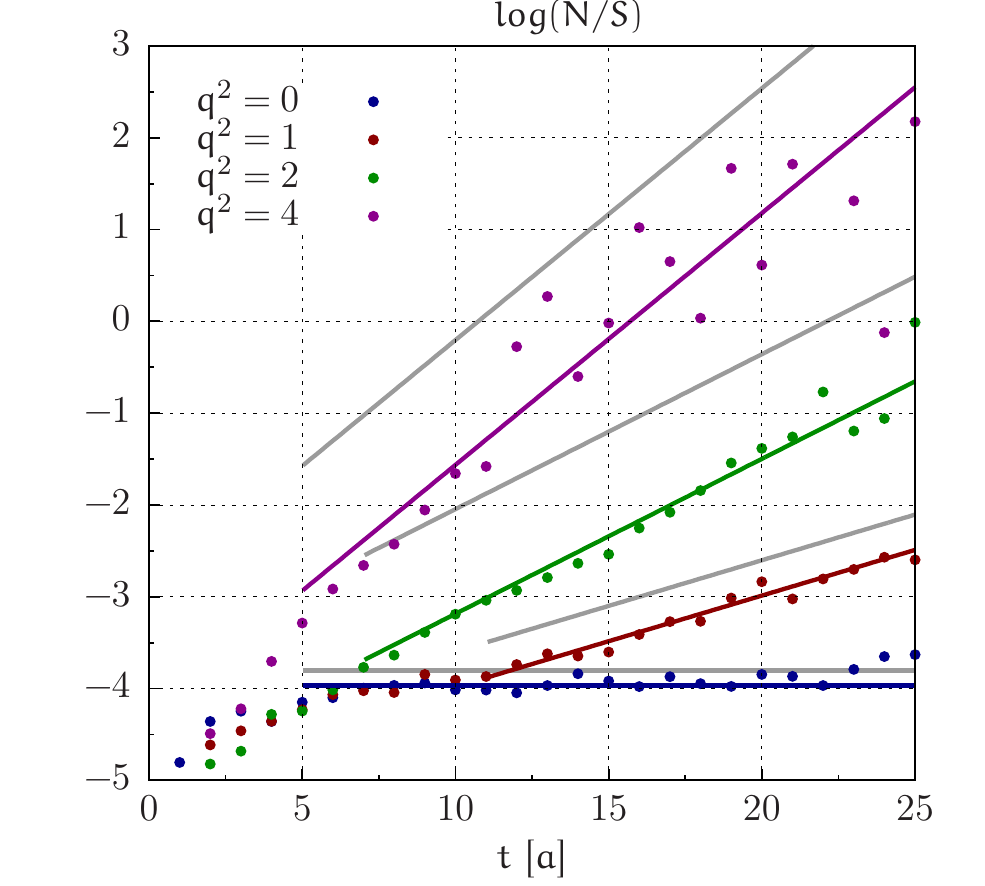}
\caption{\small Noise-to-signal ratio for a pion for different $\bf{q}^2$ plotted against the expected asymptotic 
behaviour for anisotropic smearing. The grey lines show the corresponding behaviour for spherical smearing (the 
corresponding spherical data points are omitted).}
\label{fig:RNSasy}
\end{center}
\end{figure}

\begin{figure}
\centering
\includegraphics[width=0.9\linewidth]{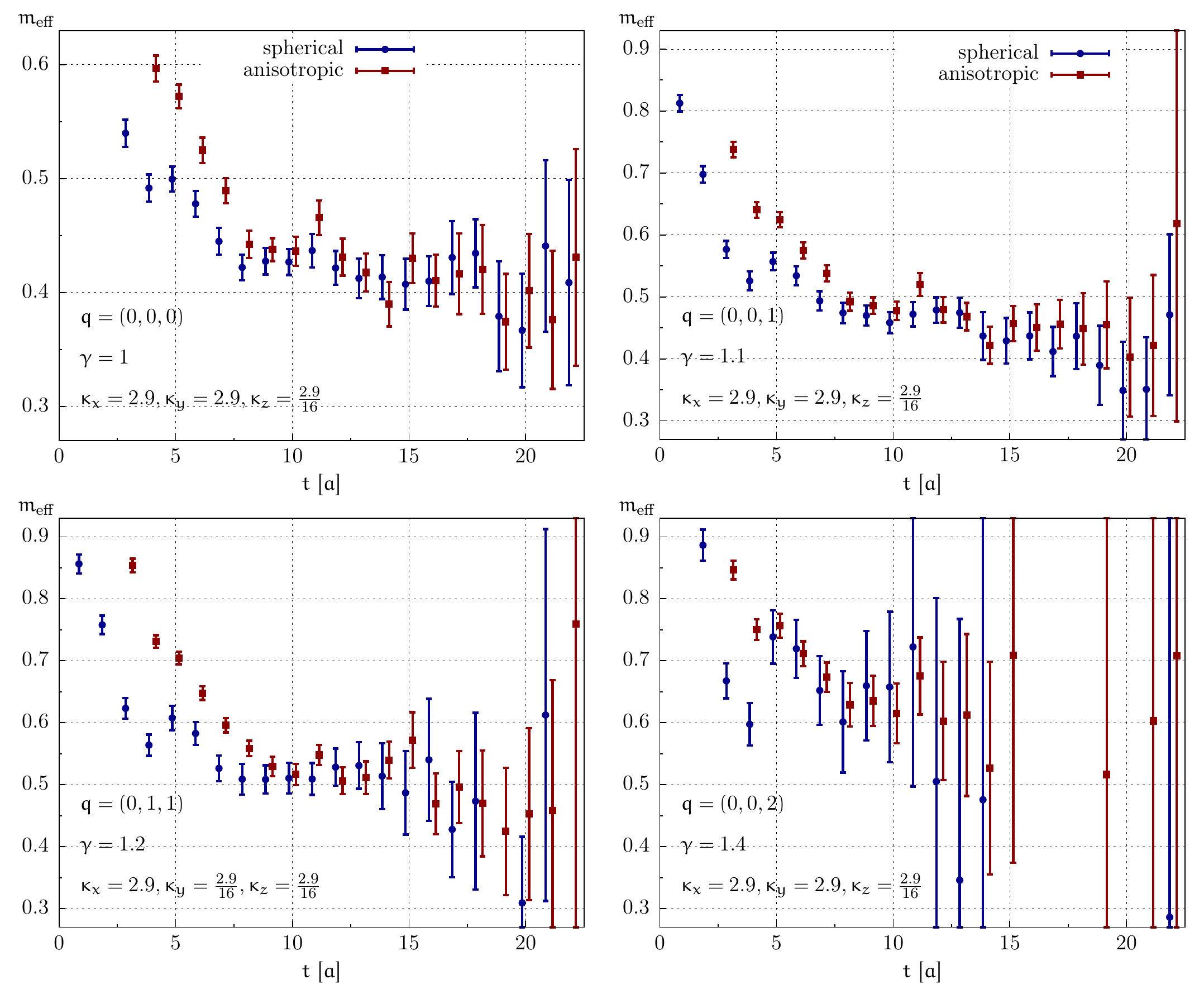}
\caption{\small Effective energies for the nucleon for different values of ${\bf q^2}$ using spherical (blue) and anisotropic (red) smearing.}
\label{fig:Meff_Nucl}
\end{figure}

The results demonstrate that anisotropic smearing offers a clear improvement for non-zero momenta, which 
becomes more pronounced for larger boosting factors. This is further demonstrated in figure~\ref{fig:RNSasy}.
We investigated the effects of adjusting the anisotropy in different directions by multiplying the appropriate
components of $\boldsymbol{\kappa}$ with a factor $1/N$, where $N=2$, 4, 8, 16 and 32. We found that by $N=16$ the improvement in the 
statistical noise had saturated and any further improvement was accompanied by a slightly worsened 
overlap with the ground state. We explicitly checked that squeezing in directions orthogonal to those with 
momenta provided no gain. The same techniques were applied to the nucleon (figure~\ref{fig:Meff_Nucl}), 
however, due to the worse initial signal and larger mass, the nucleon only attains relatively small boosting 
factors before the signal is lost in noise and so the effect is less pronounced.

\begin{figure}
\begin{center}
\includegraphics[width=0.9\linewidth]{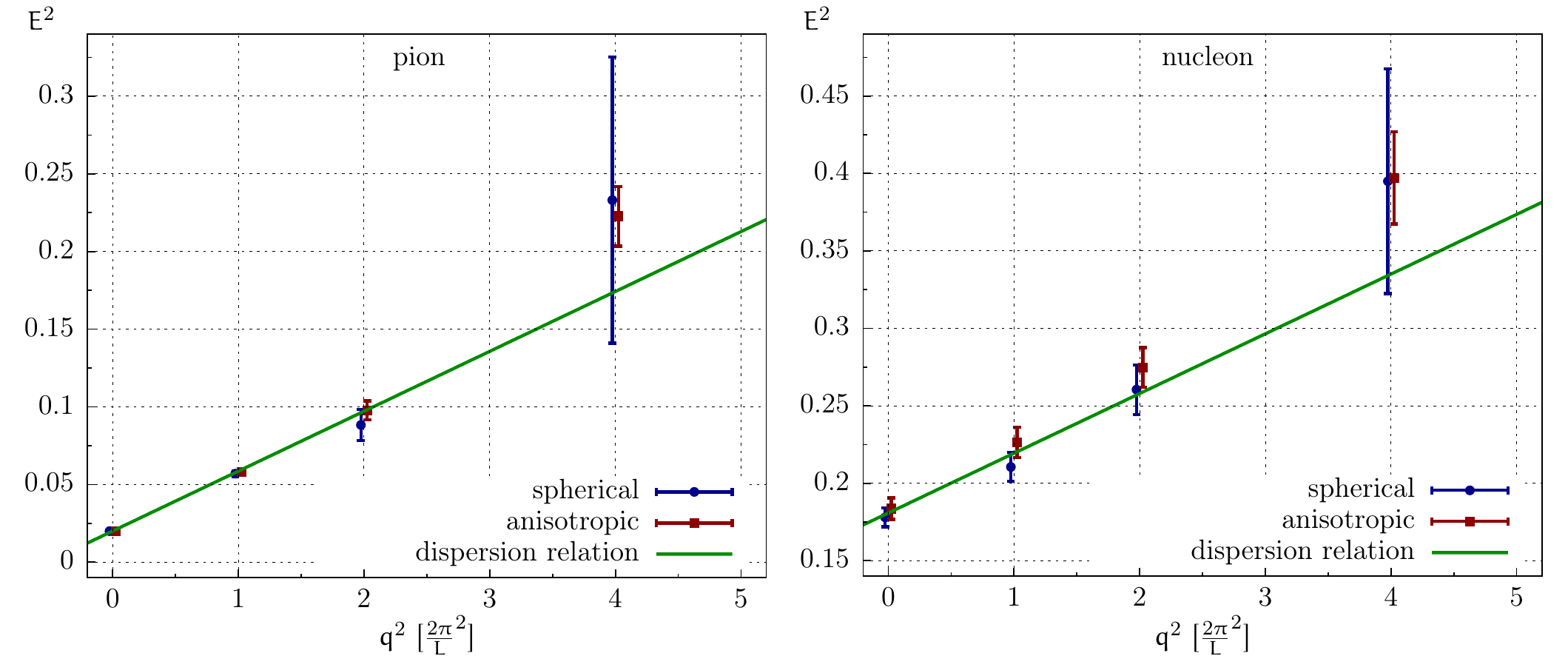}
\caption{\small The dispersion relation plotted against the extracted effective energies for spherical (blue) and anisotropic (red) 
smearing for the pion (left panel) and nucleon (right panel). The solid
    line denotes the continuum dispersion relation.}
\label{fig:Disp}
\end{center}
\end{figure}

Figure~\ref{fig:Disp} compares the extracted results for the spherical and anisotropic smearing to the 
continuum dispersion relation for both the pion and nucleon, further demonstrating the potential statistical gain.
The extracted values were obtained via a $\chi^2$ per d.o.f. minimisation of simple plateau fits to the data 
in figures~\ref{fig:Meff_Pion} and \ref{fig:Meff_Nucl}, and the same fitranges are used for both spherical 
and anisotropic smearing.

\section{Conclusion and outlook}

We have implemented a generalisation of Gaussian/Wuppertal smearing which allows the smearing to be optimised 
to account for a hadron's kinematics, which for a boosted hadron correspond to an anisotropic wavefunction. 
Our data demonstrates that this provides a significant improvement in the noise-to-signal ratio for heavily 
boosted hadrons whilst maintaining a good overlap with the ground state. The results are more pronounced for the pion
due to its large boosting factor at small values of the lattice momenta. 
Our results show that the improvement is more dependent on $\gamma$ than on the mass of the hadron, and as such we expect
this prescription to be more effective as the chiral limit is approached. Important applications for this method
could include the determination of mesonic and baryonic form factors, such as for $B\rightarrow \pi l\nu$ or the 
nucleon form factors $G_E$ and $G_M$, for which we presented preliminary results at this conference \cite{talk}.
A preliminary study of the nucleon three-point function used 
in the calculations of the nucleon form factors show that a similar gain could be achieved. 
We are currently extending this study.
In principle, anisotropic smearing requires a different smearing for each choice of momentum and so the largest gain would 
likely be achieved though the use of all-to-all \cite{Foley:2005ac,Endress:2011jc} stochastic propagators to reduce the number of inversions 
of the Dirac operator and thus the computational cost.   

A published account of this work may be found in~\cite{DellaMorte:2012xc}. We note that a similar approach was applied to the
nucleon channel in~\cite{Roberts:2012tp}, which is consistent with our findings.

\end{document}